\documentclass[12pt,preprint]{aastex}

\begin{document}

%\input def.tex
% Macro definitions for the z=5.7 candidates paper for LALA
%
\newcommand{\lya}{Lyman-$\alpha$}
\newcommand{\eqw}{\hbox{EW}}
\def\erg{\hbox{erg}}
\def\cm{\hbox{cm}}
\def\sec{\hbox{s}}
\def\f17{f_{17}}
\def\Mpc{\hbox{Mpc}}
\def\nm{\hbox{nm}}
\def\km{\hbox{km}}
\def\kms{\hbox{km s$^{-1}$}}
\def\year{\hbox{yr}}
\def\deg{\hbox{deg}}
\def\arcsec{\hbox{arcsec}}
\def\microJy{\mu\hbox{Jy}}
\def\zre{z_r}
\def\fesc{f_{\rm esc}}

\def\ergcm2s{\ifmmode {\rm\,erg\,cm^{-2}\,s^{-1}}\else
                ${\rm\,ergs\,cm^{-2}\,s^{-1}}$\fi}
\def\ergsec{\ifmmode {\rm\,erg\,s^{-1}}\else
                ${\rm\,ergs\,s^{-1}}$\fi}
\def\kmsMpc{\ifmmode {\rm\,km\,s^{-1}\,Mpc^{-1}}\else
                ${\rm\,km\,s^{-1}\,Mpc^{-1}}$\fi}
\def\nv{\ion{N}{5} $\lambda$1240}
\def\civ{\ion{C}{4} $\lambda$1549}
\def\oii{[\ion{O}{2}] $\lambda$3727}
\def\oiipair{[\ion{O}{2}] $\lambda \lambda$3726,3729}
\def\oiii{[\ion{O}{3}] $\lambda$5007}
\def\oiiipair{[\ion{O}{3}] $\lambda \lambda$4959,5007}

\title{Lyman-$\alpha$ Emitters at Redshift z=5.7}

\author{James E. Rhoads \altaffilmark{1,4} \and
Sangeeta Malhotra \altaffilmark{2,3,4}}

\begin{abstract}
\lya\ galaxies at high redshifts offer a powerful probe of both the
formation of galaxies and the reionization of the intergalactic
medium.  \lya\ line emission is an efficient tool for identifying
young galaxies at high redshift, because it is strong in systems with
young stars and little or no dust --- properties expected in galaxies
undergoing their first burst of star-formation.
\lya\ galaxies also provide a robust test of the reionization epoch that
is independent of Gunn-Peterson trough observations in quasar spectra
and is better able to distinguish line center optical depths $\tau
\sim 5$ from $\tau \sim 10^5$.  This is because neutral gas
scatters \lya\ photons, dramatically ``blurring'' images of \lya\ galaxies
embedded in a neutral intergalactic medium and rendering them undetectable.

We present a photometrically selected sample of redshift $z \approx
5.7$ \lya\ emitters derived from the Large Area Lyman Alpha survey.
The presence of these low-luminosity \lya\ sources immediately implies
that the reionization redshift $\zre > 5.7$.  Comparing these objects
to our earlier $z\approx 4.5$ sample, we find that the number of $z
\approx 5.7$ emitters at fixed line luminosity marginally exceeds the
no-evolution expectation, but falls well short of published model
predictions.  The equivalent width distribution is similar at the two
redshifts.  The large equivalent widths of the \lya\ line indicate
young galaxies undergoing their first star formation.
\end{abstract}

\altaffiltext{1}{Space Telescope Science Institute, 3700 San Martin Drive, Baltimore, MD 21218} 
\altaffiltext{2}{Johns Hopkins University, Charles and 34th Street, Bloomberg center, Baltimore, MD 21218}
\altaffiltext{3}{Hubble Fellow} 
\altaffiltext{4}{Visiting Astronomer, Kitt Peak National Observatory,
National Optical Astronomy Observatory, which is operated by the
Association of Universities for Research in Astronomy, Inc. (AURA)
under cooperative agreement with the National Science Foundation.}

\keywords{galaxies: general --- galaxies: evolution --- galaxies: formation 
--- galaxies: statistics --- cosmology: observations --- early universe}

\section{Introduction}

Recent discoveries of high redshift objects are steadily taking us
closer to determining the epoch of galaxy formation (Dey et al. 1998;
Spinrad et al. 1998; Weyman et al. 1998; Hu, McMahon, \& Cowie 1999;
van Breugel et al. 1999; Fan et al. 2001), if such a well-defined time
period exists. Observations of neutral hydrogen absorption troughs
(the Gunn-Peterson effect) seem to indicate that intergalactic
hydrogen was reionized and hence that the ionizing UV ``turned on''
between $z\approx 6.3$ and $z\approx 5.7$ (Becker et al. 2001,
Djgorvski et al. 2001).
%  The difference between reionization redshifts
%  for hydrogen ($\zre \sim 6$) and for HeII ($\zre \sim 3$) links the
%  hydrogen reionization to galaxy formation: If hydrogen were reionized
%  by quasars, which have harder spectra, HeII would be ionized to HeIII
%  at about the same time (e.g., Madau \& Meiksin 1994; Kriss et al 2001).

A more direct probe of galaxy formation is to determine the
number counts and redshift evolution of nascent galaxies.
The Large Area Lyman Alpha (LALA) survey (Rhoads et
al. 2000) has been designed and executed to detect large enough
samples of high-redshift emitters to be statistically useful. Strong
\lya\ emission, shown by large equivalent widths in the
line, indicates that these are young galaxies (Malhotra et
al. 2001). Since large scale structure can influence
galaxy counts even at high redshifts (Steidel et al. 1998, Adelberger
et al. 1998), large area surveys are vital to study redshift evolution.
Narrow-band surveys maximize sensitivity and solid angle,
and allow us to push for higher redshift objects by designing
narrow-bands to avoid bright sky lines.

\lya\ galaxies also offer a particularly direct and robust test of
the reionization epoch (Haiman \& Spaans 1999).  Before reionization,
\lya\ emitting galaxies of normal luminosity are effectively hidden
from view by the scattering ``fog'' of the neutral IGM
(Miralda-Escud\'{e} 1998; Rybicki \& Loeb 1999).  Thus, the
reionization redshift should be marked by a sharp decrease in the
number counts of faint \lya\ emitters.  This test is independent of
the Gunn-Peterson trough observations in quasar spectra.  Moreover,
the Gunn-Peterson test has difficulty distinguishing opaque but modest
optical depths $\tau \sim 5$ from the much larger obscuration $\tau
\sim 10^5$ expected in a neutral IGM.
\lya\ galaxy counts can resolve this ambiguity.  The red wing of the
\lya\ line from a star forming galaxy sees only the red damping wing
of intergalactic \lya\ absorption, and \lya\ galaxies therefore remain
detectable up to line-center optical depths $\tau \sim 10^4$.

In this paper we report on a population of candidate $z\approx 5.7$
\lya\ emitters discovered in the second phase of LALA survey and
compare their properties to the z=4.5 sample (Rhoads et al. 2001).
Section~\ref{obs} describes our observations and data reduction. In
section 3 we present the candidates and their properties. Section 4
explores the implications of these discoveries for the epoch of galaxy
formation.

\section{Observations and data analysis}  \label{obs}

 The Large Area Lyman Alpha survey began in spring 1998 as a
narrowband search for $z \approx 4.5$ \lya\ emitting galaxies.  The
project was designed to exploit the high survey efficiency offered by
the 36' field of the new CCD Mosaic Camera at the Kitt Peak National
Observatory 4m Mayall Telescope.  The core of the LALA survey is $0.72
\,\Box^\circ$ in two fields centered at 14:25:57 +35:32 (2000.0) (the
Bo\"{o}tes field) and 02:05:20 -04:55 (2000.0) (the Cetus field).

In April 2000, we extended the Bo\"{o}tes field survey to $z=5.7$
using custom-built narrowband filters having central wavelengths
$\lambda_c = 815, 823\nm$, full width at half maximum (FWHM)
transmission $7.5\nm$, and peak throughput $\sim 90\%$.
The wavelengths were chosen to fall in a gap in the night sky
emission line spectrum, thus minimizing sky noise.
The filter bandpass deteriorated below these specifications outside a
central circle of $\sim 30'$ diameter.  This resulted in substantially
degraded throughput in the outer parts of our images, which we therefore
excluded from our analysis.

The $z=5.7$ observing run was four nights, beginning with the night of
10--11 April 2000 (i.e., 2000 April 11 UT).  The weather was generally
clear.  Photometric stability was checked by measuring
the fluxes of selected stars on all science exposures.  This showed $<
11\%$ variation in throughput for the entire run, mostly due to
airmass-dependent extinction.
% The variation is 6% for the 823 filter.
% it is 11% for 815.

We obtained 9.85 hours of data in 40 exposures through the
815 nm filter, and 9 hours in 36 exposures through the 823 nm filter.
Exposures were spatially dithered with characteristic (RMS) offsets of
$\sim 50''$.  Seeing ranged from $0.75''$ to $2.25''$ (median
$0.93''$) in the 815 nm filter, and $0.78''$ to $1.25''$ (median
$1.01''$) in the 823 nm filter.  Median narrowband sky brightnesses
were $27 \microJy \, \arcsec^{-2}$ at 815 nm and
$21 \microJy \, \arcsec^{-2}$ at 823 nm. We also obtained a deep (4
hour) image in a Sloan Digital Sky Survey z' filter during this run.

Data reduction was done in IRAF following the methods used for
the $z=4.5$ sources (Rhoads et al 2000).
%First, crosstalk between
%chip pairs that share readout electronics was removed.  Next, overscan
%corrections, bias frames, and flatfield corrections were applied.
%Dome flats are sufficient to correct cosmetically for spatial
%variations in the filter throughput (see above).  A ghost image of the
%telescope pupil was then modeled and subtracted from each frame.
%Finally, second order flat fielding problems were removed using a
%heavily smoothed super-sky flat constructed from the object frames,
%and a polynomial surface was fitted to blank sky regions on each chip
%and subtracted in order to remove any remaining sky gradients.  Cosmic
%rays in individual exposures were identified and flagged using the
%algorithm of Rhoads (2000). Each image was then placed on a world
%coordinate system using USNO-A catalog stars in the field.  Satellite
%trails and a few internal reflections due to bright stars were flagged
%by hand for exclusion from the final stacks.  Finally, the images were
%combined using IRAF MSCRED tasks (Valdes \& Tody 1998; Valdes 1998) to
%correct geometric distortions in the images.  Weights for the final
%stack were determined using the ATTWEIGHT code of Fischer \& Kochanski
%(1994), which accounts for sky level, transparency, and seeing to
%optimize the signal to noise level for compact sources in the final
%image.
%
The final stacks have point spread functions (PSF) with directly
measured full widths at half maximum of $0.88''$ ($815\nm$) and
$0.97''$ ($823\nm$).  The PSF wings are higher than Gaussians of the
same FWHM, since the images are sums of many frames with varying
seeing. Catalogs were generated using SExtractor (Bertin
\& Arnouts 1996).  Fluxes were measured in $2.32''$ (9 pixel) diameter
apertures, and colors were obtained using matched $2.32''$ apertures
in registered images.  We also measured fluxes in earlier
broad band images from the NOAO Deep Wide-Field Survey (B$_W$, R, and
I bands) (Jannuzi \& Dey 1999) and broad and narrow band images from
previous LALA runs (V band and four $\lambda \approx 6600$\AA\ narrow
bands).

We select $z \approx 5.7$ candidates using the following criteria:
(1) Secure detection in a narrowband filter ($> 5 \sigma$).
(2) A strong narrowband excess (narrow $-$ broad color $< -0.75$ magnitude,
corresponding to a factor of $\approx 2$ in flux density)
that is securely measured (the flux density in the narrow band should
exceed that in the broad band at the $4 \sigma$ level).
(3) No flux at wavelengths shorter than the expected Lyman break
wavelength (the object should remain undetected in B$_W$ and V band fluxes
at the $2 \sigma$ level or below).
The first two criteria are designed to ensure that real emission 
line objects are selected, while the third rejects most low redshift
emission line galaxies, using the blue bands as ``veto'' filters.
% Magnitudes and colors were computed using ``asinh'' magnitudes
% (Lupton, Gunn, \& Szalay 1999), which account for upper limits more
% gracefully than do traditional (purely logarithmic) magnitude scales.

\section{Survey Results}  \label{results}

\begin{figure}
\epsscale{0.99}
% \plotone{z5.7_fig_v3.ps}
\plotone{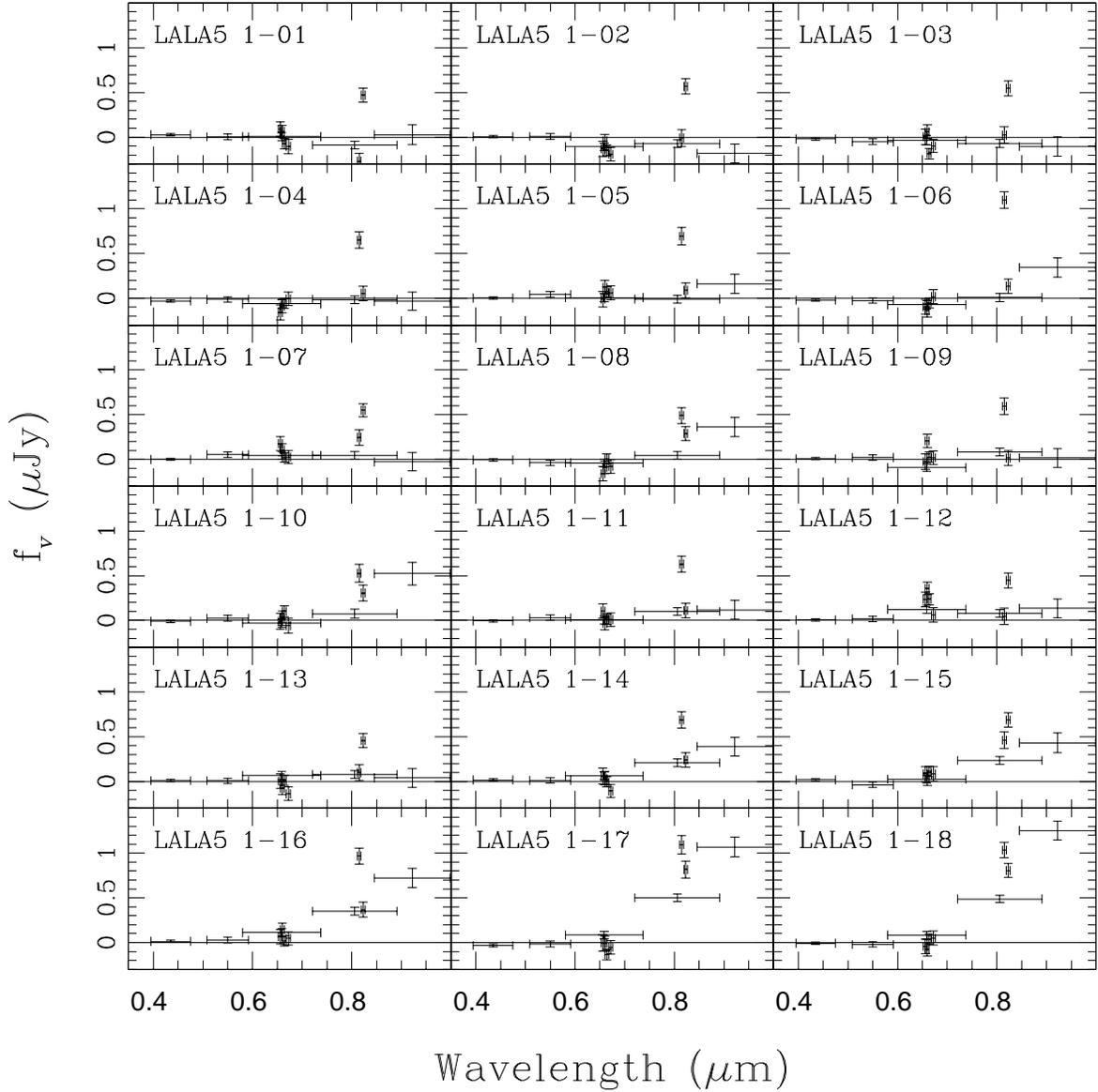}
\caption{
Photometric ``mini-spectra'' for the 18 $z \approx 5.7$ 
\lya\ emitter candidates selected using the criteria discussed in
section~\ref{obs}.
All photometry was done through a $2.32''$ diameter aperture
centered on the object position derived from the narrowband filter
containing the excess emission.  
The candidates are assigned ranked ID numbers based on 
the ratio of broad to narrow band
flux density, which is a measure of emission line strength.
This rank increases from left to right and from top to bottom.
The last five candidates (LALA5 1-14 to LALA5 1-18) have observer frame
equivalent widths below $80$\AA\ if we correct the continuum flux
for the intergalactic hydrogen absorption expected at $z\approx 5.7$.
\label{minispec}}
\end{figure}

In total, 18 objects in the two filters (11 in the $815\nm$ filter and
7 in the $823\nm$ filter) passed all selection criteria and are
considered good $z \approx 5.7$ candidates.  Photometric
``mini-spectra'' for these candidates are shown in figure~\ref{minispec}.

Intergalactic hydrogen absorption is expected to attenuate the
measured I band flux used for candidate selection.  
The formulation of Madau (1995) gives
correction factors of $2$ for $z=5.70$ and $2.2$ for $z=5.77$.
We apply these corrections to the I band flux before calculating the
equivalent widths of the emission lines from our broad and narrow band
photometry.  These corrections are comparable to the factor of 2
difference in flux density required for selection.  We therefore apply
another cut to our candidate lists, removing the five candidates whose
observer frame equivalent widths fall below $80$\AA\ after the
correction for hydrogen absorption. 
We expect these objects to be either high redshift {\it
or\/} strong emission line objects, but they need not be both.
Because intergalactic hydrogen absorption can also remove the bluest
$\sim 50\%$ of the \lya\ line flux, our equivalent width corrections
and the resulting candidate list cut are both quite conservative.

\section{Discussion and Conclusions} \label{discuss}

We now turn to the implications of our survey for galaxy formation and
cosmology. First, let us compare the statistics of $z\approx 5.7$
sources to the $z\approx 4.5$ \lya\ emitters.  There are 13 sources at
$z\approx 5.7$ and 156 at $z\approx 4.5$.  However, a minority of our
$z\approx 4.5$ are bright enough to be detectable at $z\approx 5.7$.

Using a $\Omega_m = 0.3$, $\Omega_\Lambda = 0.7$ cosmology and accounting
for the difference in solid angle between the two
redshifts\footnote{The solid angle for the higher redshift is approximately
a factor of two smaller due to the degradation of the bandpass near the
filter edges; see section~\ref{obs}.}, the comoving survey volume at
$z\approx 5.7$ is $31\%$ that at $z\approx 4.5$ ($16\%$ in each filter).
The $5\sigma$ flux thresholds used to identify candidates at $z\approx 5.7$
were $0.455 \microJy$ at $815\nm$ and $0.405 \microJy$ at $823\nm$.  The 
distance-corrected equivalent thresholds at $z\approx 4.5$
will be larger by a factor of $1.34^2 = 1.8$ for the above cosmology. 
There are then $15$ $z \approx 4.5$
candidates that could be detected in our $815\nm$ data, and $26$
that could be detected in our $823\nm$ data.  Thus, based on the
$z\approx 4.5$ survey and a no-evolution assumption, we would expect
to find $2.3$ sources in the $815\nm$ filter, and $4.1$ in the
$823\nm$ filter.  We actually identify $7$ and $6$ respectively.

This excess in the observed $z\approx 5.7$ candidate counts is
moderately significant (at the 99\% level assuming Poisson source
counts).  The important difference is in the $815\nm$ counts;
the $823\nm$ filter candidate counts are consistent with the
na\"{\i}ve scaling from $z\approx 4.5$.  We now examine several
possible explanations for this effect.

% Evolution
First, \lya\ emitters at $z\approx 5.7$ may be more numerous and/or more
luminous than those at $z\approx 4.5$.  Such effects are reasonable:
The \lya\ line is resonantly scattered and can be strongly quenched by
dust absorption.  Since dust is formed from stellar nucleosynthesis
products, we expect galaxies at earlier epochs to be (on average) less
chemically evolved and less dusty.  More luminous \lya\ lines may be a
natural byproduct, provided that the effects of chemical evolution are
not overwhelmed by the trend of decreasing halo mass at higher
redshift. Indeed, under the \lya\ emitter population models of Haiman
\& Spaans (1999), evolutionary effects are so strong that the
total surface density per unit redshift above a fixed line flux $\sim
1.5 \times 10^{-17} \ergcm2s$ is approximately independent of redshift
for $3 \la z \la 6$.  Our results show a factor $\sim 7$ decrease in
the surface density of \lya\ emitter candidates from $z=4.5$ to
$z=5.7$ (in contrast to an expected factor $< 2$ from the Haiman \&
Spaans models).  The LALA data is thus more nearly consistent with a
no-evolution model than with the baseline model from Haiman \& Spaans
(1999).

% Large scale structure...
In addition, large scale structure can significantly affect the number
counts of objects in single fields.  Our $36'$ field makes LALA results
comparatively robust to such effects, but
a factor of $2$ difference in source density may still be consistent
with cosmic variance.

Finally, our candidate sample will contain some residual contamination
by foreground objects.  The numbers of such objects should be
comparatively small, given the high-redshift consistency checks
described in sections~\ref{obs} and~\ref{results}. 
Foreground emission line galaxies are the most serious concern.
Our primary defense against such galaxies is the
requirement that a good candidate be undetected in deep broadband
images with $\lambda < (1+z) \times 912$\AA.  This imposes an
approximate minimum observer frame equivalent width for any
interloper, which we estimate using the minimum narrow band flux $5
\sigma_n$ and maximum broad band flux $2 \sigma_b$.  (Here $\sigma_n$
and $\sigma_b$ are the photometric errors in flux density units for
the narrow and broad blue filters.)  The corresponding equivalent
width is then $\approx (f_n / f_b - 1) \times \Delta\lambda_n = \left[
5 \sigma_n / (2 \sigma_b) - 1 \right] \times \Delta\lambda_n \sim
1100$\AA, where we assume that the line lies outside the broad band
filter.  The redshifts of likely foreground objects are relatively low
($0.24 \la z \la 1.21$), and the corresponding emission frame
equivalent widths therefore remain large ($\ga 500$\AA).  While
compact, narrow emission line galaxies can attain these equivalent
widths, they are very rare (e.g., Hogg et al 1998).

Cool stars can also enter narrowband emission line samples because
molecular absorption in their atmospheres produces spectral
features with widths of a few $\nm$ and amplitudes several tens of
percent.  However, in the absence of true emission lines, cool stars
will show detectable continuum at $\ga 1/2$ of the
narrowband flux density, which is also expected to be quite red, so
that our SDSS z' filter should clearly show such stars.  Most of our
candidates having strong z' detections were discarded from our sample
by the last cut applied (see section~\ref{results}).  Only two of the
13 remaining sources, LALA5 1-08 and LALA5 1-10, have ``minispectra''
that plausibly resemble cool stars.

Because our narrow and broad band filters were obtained at different
times, variable objects may also enter the sample at a low level.  The
relevant objects would have to have $I \la 24.3$ during the
narrow band observations (2000 April 11--14 UT), and $I>25$ during
the I band observation (1999 March 27 UT).  (In
fact, our requirement on the blue flux imposes additional limits.)
The largest known source of such variable objects would be high
redshift supernovae.  However, these would show roughly
equal brightness in both $z\approx 5.7$ narrow bands and the SDSS z'
image, since their characteristic variability timescale is several
days.  Again, only LALA5 1-08 and LALA5 1-10 are
consistent with this expectation.  Sources with faster variability,
such as orphan GRB afterglows (Rhoads 1997), are very rare and we
would expect $\ll 1$ such source in our sample (e.g., Dalal et al 2001).

We compared the equivalent width distribution of the $z=4.5$ and $5.7$
samples. Very low and high equivalent width objects are missing in
z=5.7 sample compared to z=4.5 sample. 
Using a Kolmogorov-Smirnov test we find
a 20\% chance that the two samples are
drawn from the same parent distribution. This test is only
marginally conclusive, mostly because of the small number of $z=5.7$
objects.  The consistency of the present samples is nonetheless
further evidence that we are seeing the same class of object at
$z=5.7$ as at $z=4.5$.

%\begin{figure}
%% \plotone{ks.ps}
%\plotone{f2.eps}
%\caption{
%Cumulative equivalent width histograms for $z=4.5$ (solid line) and
%$z=5.7$ (dotted line) candidate \lya\ emitters from the LALA survey.
%The last bin includes all sources for which the data yield lower
%limits to the equivalent width.  A Kolmogorov-Smirnov test shows
%that the two distributions are consistent (see text).
%\label{ks_fig}}
%\end{figure}

In summary, we have a sample of $13$ $z\approx 5.7$ \lya\
emitting galaxy candidates.  Only two of these have the
apparent spectral energy distributions expected of stars or variable
objects.  The upper limits on blue flux rule out all but the highest
equivalent width foreground sources, which are too rare to account
for the entire sample.  The equivalent width distributions are 
consistent at $z\approx 5.7$ and $4.5$.  Thus, the simplest
explanation of the data is that we are really seeing $z\approx 5.7$
\lya\ emitters.

% Expectations under assumption that reionization occurred at z=5.x
We now draw one very robust and important conclusion from our sample.
The epoch of reionization is a major landmark in the evolution of
the universe.  Recently, the Gunn-Peterson trough has been reported
in spectra of two high redshift quasars: SDSSpJ 103027.10+052455.0
at $z = 6.28$ (Becker et al 2001), and SDSS 1044-0125 at $z = 5.73$
(Djorgovski et al 2001).  These authors have interpreted their spectra
as evidence that reionization was incomplete at these redshifts.

If the universe was neutral at redshift $\zre < 5.67$, \lya\ photons from
sources at higher redshift would be resonantly scattered in the
neutral intergalactic medium (IGM).  This effectively erases the \lya\
line from view for a survey like LALA. Thus, our $z\approx 5.7$ sample
implies that the reionization redshift lies at $\zre>5.8$ along the line
of sight to the LALA Bo\"{o}tes field.  This argument relies merely on
the {\it presence\/} of low-luminosity \lya\ sources, and is
independent of their exact number.

Two possible loopholes exist, but both
are readily closed.  First, in the absence of dust, the \lya\ photons
eventually scatter far enough into the line wing to propagate freely.
However, this process spreads the line emission greatly in both area
and frequency, resulting in a very large ``photosphere'' and a
correspondingly low surface brightness (Loeb \& Rybicki 1999).  All of
the sources reported in this paper are compact in the narrow band
image (i.e., essentially unresolved at our resolution limit of $\la
1''$) and so morphologically inconsistent with a source embedded in a
neutral IGM.

Second, a source with a sufficient ionizing flux will generate
its own Stromgren sphere in the IGM.  If this sphere is large enough,
\lya\ photons will redshift by $\ga 1000 \kms$ before reaching the
surrounding neutral gas, and will thus avoid resonant scattering.
However, this effect requires sources either more luminous or older than
typical LALA objects.  We assume that a fraction $\fesc$ of ionizing
photons escape a galaxy to ionize its Stromgren sphere, and that $2/3$
of the remainder result in \lya\ photons, and that $H_0 = 70 \kmsMpc$,
$\Omega_m=0.3$, $\Omega_\Lambda = 0.7$, and $\Omega_b =0.05$.  We then
require an ionized bubble radius $r_s \ga 1.2 \Mpc$, which gives a
Gunn-Peterson effect optical depth $\tau \le 1$ at emitted line center
(see also Loeb \& Rybicki 1999).  Ignoring recombinations, $r_s \ga
1.2 \Mpc$ requires a source with $L_{43} t_8 \fesc/(1-\fesc) \ga 5$,
where $L_{Ly-\alpha} = 10^{43} L_{43} \ergsec$ and the source is $10^8
t_8$ years old.  LALA $z\approx 5.7$ sources have $L_{43} \approx
0.7$.  Their median rest frame equivalent width of $\ge 80$\AA\ 
requires $\fesc \la 0.2$ for any reasonable initial mass
function\footnote{Continuous star formation with IMF slope
$\alpha=2.35$ (i.e., Salpeter), upper mass cutoff $120 M_\odot$,
metallicity $Z = Z_\odot/20$, and $t_8 \ga 0.3$ gives $EW \approx 95
(1-\fesc)$\AA.  Substantially larger EW requires $\alpha \ll 2$.}, unless
$t_8 \la 0.3$ (Malhotra et al 2001; Kudritzki et al 1999).  The age of
the universe at $z=5.7$ requires $t_8 < 10$.  Thus, $L_{43} t_8
\fesc/(1-\fesc) \la 2$, and \lya\ radiation from these sources would
indeed suffer resonant scattering if the IGM were neutral.

We are thus confident that the compact, low-luminosity \lya\ emitters
we observe at $z \approx 5.7$ are in a mostly ionized universe.
This method offers a test for reionization 
independent of the spectroscopic search for Gunn-Peterson troughs.
Moreover, it is better able to distinguish line center optical depths
$\tau \sim 10^5$ (indicating a neutral IGM) from much smaller but still
essentially opaque optical depths $\tau \sim 10$ (indicating \lya\
forest absorption), because the \lya\ forest will in general absorb
only the blue side of the \lya\ emission line, while a neutral IGM
will absorb the entire line.
For an intrinsic line width of $\sim 100 \Delta v_{100} \kms$, we 
would still expect to see some \lya\ flux through an IGM with a
(homogeneously mixed) neutral fraction of $\la 0.1 \Delta v_{100}
\left((1+z)/6.7\right)^{-3/2}$, while the Gunn-Peterson trough is
optically thick ($\tau \ge 5$) for a neutral fraction as small as
$\sim 4 \times 10^{-5} \left((1+z)/6.7\right)^{-3/2}$.

To conclude, we have demonstrated that \lya\ emitters are found at
$z\approx 5.7$ in densities comparable to those at $z\approx 4.5$ at
similar line luminosities.  It follows that the reionization redshift
was $\zre > 5.8$, at least along this line of sight.  Tentatively, it
appears that the \lya\ source density at equal luminosity rises
from $z\approx 4.5$ to $z\approx 5.7$. We have also obtained pilot
observations in a $z\approx 6.6$ narrowband filter.  This redshift
lies beyond the recently reported Gunn-Peterson effect quasars, and
will ultimately allow us to bracket the reionization epoch through
\lya\ source counts.

\acknowledgements 
This work made use of images provided by the NOAO Deep Wide-Field
Survey (NDWFS; Jannuzi and Dey 1999), which is supported by the National
Optical Astronomy Observatory (NOAO).  NOAO is operated by AURA,
Inc., under a cooperative agreement with the National Science
Foundation.
We thank Buell Jannuzi, Arjun Dey, and the rest of the NDWFS team for
making their images public; and Richard Green and Jim De Veny for
their support of the LALA survey.
JER's research is supported by an Institute Fellowship at The Space
Telescope Science Institute (STScI).
SM's research funding is provided by NASA through Hubble Fellowship
grant \# HF-01111.01-98A from STScI.
STScI is operated by the Association of Universities for Research in
Astronomy, Inc., under NASA contract NAS5-26555.

\end{document}